# Design approaches in technology enhanced learning

**Yishay Mor and Niall Winters**

London Knowledge Lab,

Institute of Education, University of London

{y.mor, n.winters}@ioe.ac.uk

## Abstract

Design is a critical to the successful development of any interactive learning environment (ILE). Moreover, in technology enhanced learning (TEL), the design process requires input from many diverse areas of expertise. As such, anyone undertaking tool development is required to directly address the design challenge from multiple perspectives. We provide a motivation and rationale for design approaches for learning technologies that draws upon Simon's seminal proposition of Design Science (Simon, 1969). We then review the application of Design Experiments (Brown, 1992) and Design Patterns (Alexander et al., 1977) and argue that a patterns approach has the potential to address many of the critical challenges faced by learning technologists.

## Keywords

# Introduction

The design of a technology enhanced learning (TEL) tool, and specifically of Interactive Learning Environments (ILEs), is a major challenge. This is because it must address issues ranging from learning theory to software engineering. Developers face fundamental challenges in building tools to adequately address the issues raised during the design process. However, understanding and resolving each of the requirements and the tensions between participants has long been recognized as fundamental to any tool's success. In this paper, we present a thematic review of design approaches in TEL, highlight some of the key challenges and suggest that a design pattern approach may offer a way forward.

We begin with a reflection on the historical foundations of design research, focusing on the work of Herbert Simon. Revisiting Simon's work we found it just as perceptive today as on the day it was published. This leads us to explore contemporary trends in design research in education, and the issues which they raise. Of particular interest is the emerging methodology of participatory design. Our next section explores the idea of design patterns, from its roots in theory of planning to current day educational research. We conclude with several open issues for future investigation.

# Foundations: Learning as a Design science

Design approaches in technology-enhanced learning generally and Interactive Learning Environments in particular, are strongly influenced by the seminal work of Simon (1969), who was the first to refer to design as a science. Simon distinguishes between the natural sciences and the sciences of the artificial, challenging the view of the latter as 'practical' science or 'vocational arts'. At the core of the study of the artificial, Simon places the science of design. In his words, "*everyone designs who devises courses of action aimed at changing existing situations into desired ones*" (Simon, 1969, p 129).

Simon's (1969) concept of design science entails more than a shift in the subject of study. It calls for a change in scientific agenda. Whereas natural science is concerned with what *is*, design science asks what *ought to be*. A rocket scientist will not be satisfied with describing and understanding existing engines. She will strive to identify the principles which will allow us to create better engines. When shifting our focus from engineering to social subjects – such as learning mathematics – the *value* aspect of design sciences becomes salient. Arguably, neurobiology and psychology investigate learning from the perspective of a natural science, while the science of education takes a design stance. The former are concerned with how humans learn, whereas the later asks how they *ought to learn*. The first may claim to be value neutral and objective, but the questions of education, by their imperative nature, are evidently derived from the observers' (often implicit) ethical, social and community agenda.

The second implication that Simon (1969) draws from the imperative nature of design science regards the method of problem decomposition. All the sciences proceed, to an extent, by decomposing complex problems into simpler ones. Design science is interested in purpose, intent and the shaping of the world to these ends. Therefore, Simon proposes function as the appropriate axis of decomposition. Such an approach is the premise of the design patterns method. It also leads Simon to what he calls the generator-test cycle (1969, p 149) as a viable method of achieving decomposition while acknowledging the networks of interdependencies between components. The design process iteratively generates solutions and then tests them against an array of functional requirements. Taken as a method of scientific inquiry, this translates into the design based research approaches.

A third key concern of Simon's is the place of representation in science. Design science is deeply concerned with the way problems under investigation are represented in order to illuminate our capacity to solve problems. We will see how Simon's principles are threaded through the design-centric research approaches we review.

## Design based research

The last couple of decades have witnessed the growing popularity of design research as a valuable methodology for educational research (Brown, 1992; Collins, 1992; DiSessa et al, 2003; Edelson, 2002; Reeves, 2000; Collins, Joseph & Bielaczyc, 2004; Barab & Squire, 2004; Cobb et al, 2003; Gorard et al, 2004; Taylor et al., 2004; Sloane & Gorard, 2003, Béguin, 2003). Design based research is a methodology for the study of function. Often referred to as design research or design experiments, it is concerned with the design of learning processes, taking account of the involved complexities, multiple levels and contexts of educational settings. The primary aim is to develop *domain-specific theories* in order to understand the learning process. However, although design experiments are being undertaken in a limited array of settings, these are viewed as instances of a broader class. Cobb et al. (2003) identify five characteristics of design experiments:

- The purpose of design experiments is "to develop a class of theories about both the process of learning and the means that are designed to support that learning".
- Design experiments are diverse and highly interventionist in nature. As they usually involve innovation there is often a discontinuity between 'typical' situation, which could be studied naturalistically and those of the design experiment. Particular forms of learning emerge from design experiments, which the researcher can directly engage with and specify as targets of investigation.
- Design experiments always have two faces: prospective and reflective. On the prospective side, designs are implemented with a hypothesized learning process and the means of supporting it in mind, in order to expose the details of that process to scrutiny. On the reflective side, design experiments are conjecture-driven tests, often at several levels of analysis.
- Together, the prospective and reflective aspects of design experiments result in iterative design. As conjectures are generated and tested, sometimes confirmed, at others refuted, new conjectures are developed and subjected to test.
- Theories developed during the process of experiment are modest, not merely in the sense that they are concerned with domain-specific learning processes, but also because they are accountable to the activity of design.

The design element in a design study may refer to the pedagogy, the activity or the tools used. In some cases, the researchers will focus on iterative refinement of the educational design while keeping the tools fixed, in others they may highlight the tools, applying a free-flowing approach to the activities. In yet others they will aspire to achieve a coherent and comprehensive design of the activity system as a whole.

Ann Brown (1992) puts forth the two main arguments in favour of design-based educational research. The first argument is methodological. The complexity of classroom situations does not lend itself to the procedures of laboratory research. Strict control of experiments and isolated variables are unattainable. Under these circumstances, Brown (1992) advocates "switching back and forth" between classroom and laboratory contexts to gain an enriched understanding of the phenomenon under study.

The second argument is ideological, possibly even ethical. It questions the fundamental goals of educational research. To what extent are we driven by a pure quest for knowledge, and to what extent are we committed to influencing educational practice? If we see contribution to good practice as a primary goal, then the outputs of our research should have direct bearing on it. This argument echoes Simon's observation regarding the value dimension of design sciences.

Critics of this approach would argue mainly with the first argument, questioning the scientific value and lack of "evidence" of inherently irreproducible experiments. The response to this critique is twofold: first, we modestly accept the limitations of this approach. But then, it is debatable whether ostensibly scientific methods can offer any greater validity. At the same time, one needs to be as stringent and self-critical when analysing data – precisely because we do not enjoy the protection of standardized statistical tests.

A more subtle criticism of the design-based approach scrutinizes it on its own turf: does this approach live up to its commitment to offer a contribution to educational practice? On one hand, the conditions of most design experiments do not resemble those of a normal classroom, if only due to the presence of a dedicated, highly informed researcher in the class. As argued by Alan Collins:

> Typically the experiments are carried out by the people who designed some technological innovation, so that they have a vested interest in seeing that it works. (Collins, 1992, p. 24 in Issroff and Scanlon, 2002).

On the other hand, the reported data and analyses typically include case-studies and theoretical generalizations derived from them. The former can be too specific to inform practising teachers, whereas the latter are often too abstract. Furthermore, there is a fundamental difference in the nature of knowledge produced by design experiments as they aim to *explain* phenomena, while maintaining a cautious stance on what causes them. In the words of Ann Brown:

> a 'Hawthorne effect' is what I want: improved cognitive productivity under the control of the learners, eventually with minimal expense, and with a theoretical rationale for why things work (Brown, 1992, p 167).

Perhaps the most substantial remarks on design studies in education come from two of its foremost proponents and promoters. DiSessa & Cobb (2004) warn against the drift of design research away from theory. In a manner similar to Simon's advocacy of rigour, they argue that theory is critical, both from a perspective of research and of practice. Furthermore, they claim that design studies can – and should - make significant theoretical contributions by addressing the gap between theory and practice. They suggest that design research may offer ontological innovations – new constructs for describing and discussing educational phenomena. We argue that design patterns offer a potential means for methodically deriving and discussing such constructs.

Before moving on, we wish to make a short note on the third element of Simon's perspective. While rarely in a direct reference to Simon (with the notable exception of Kafai, 1995), many studies highlight the issue of representation and its importance for learning. Noss & Hoyles (1996) observe that the issue of selecting and constructing representations is key to mathematical learning, and the potentials of alternative representations have been a prevailing concern of the Constructionist tradition. Radford (2002) provides theoretical support from a socio-cultural perspective. Balacheff and Kaput (1996) provide an extensive review of ILEs for mathematics, and highlight the continuous effort to diversify representations. Arguably, the issue of representation is inherent to mathematics, and thus emerges naturally when considering the design of learning environments in this domain. Yet similar work has been done in other fields, such as physics (Simpson et al, 2005). Ainsworth et al (2002) challenge common assumptions regarding the unconditional educational utility of multiple representations, arguing that is it strongly contingent on the design of the representing world as well as the represented one, and the relationship between them.

## Design patterns

The second major approach we present is design patterns. A design pattern "describes a problem which occurs over and over again in our environment, and then describes the core of the solution to that problem, in such a way that you can use this solution a million times over, without ever doing it the same way twice" (Alexander et al, 1977, p.x). This original definition positions a pattern as a high-level specification of a method of solving a problem by a design that specifies the context of discussion, the particulars of the problem, and how these can be addressed by the designated design instruments. And in *The Timeless Way of Building* he elaborates:

> Each pattern is a three-part rule, which expresses a relation between a certain context, a problem, and a solution.
>
> As an element in the world, each pattern is a relationship between a certain context, a certain system of forces which occurs repeatedly in that context, and a certain spatial configuration which allows these forces to resolve themselves.
>
> As an element of language, a pattern is an instruction, which shows how this spatial configuration can be used, over and over again, to resolve the given system of forces, wherever the context makes it relevant.
>
> The pattern is, in short, at the same time a thing which happens in the world, and the rule which tells us how to create that thing, and when we must create it. It is both a

> process and a thing; both a description of a thing which is alive, and a description of the process which will generate that thing (Alexander, 1979, p 247).

In our view a design pattern as a semi-structured description of an expert's method for solving a recurrent problem, which includes a description of the problem itself and the context in which the method is applicable, but does not include directives which bind the solution to unique circumstances. Design patterns have the explicit aim of externalizing knowledge to allow accumulation and generalization of solutions and to *allow all members of a community or design group to participate in discussion relating to the design*. Patterns are organized into coherent systems called *pattern languages* where patterns are related to each other. Although the use of design patterns never achieved a large following among professional architects, the idea has been embraced in several other disciplines, including software engineering (Gamma et al., 1995), hypermedia (German & Cowan, 2000) and interaction design (Erickson, 2000; Borchers, 2001). The approach has also found application in educational domains including e-learning systems (Derntl & Motschnig-Pitrik, 2004) and the design of computer science courses (Bergin, 2000).

*Pattern forms as tools of analysis*

An important characteristic of a design pattern is that it has three facets: descriptive, normative, and collaborative. It is an analytic form, used to describe design situations and solutions, a meta-design tool, used to highlight key issues and dictate a valuable method of resolving them, and a communicative tool enabling different communities to discuss design issues and solutions. The tension between these three aspects is visible in Alexander's work, and in much of the literature that followed. We will touch on this issue shortly.

The original collection by Alexander et al. (1977) can arguably be positioned on the *normative* end of the scale, in the sense that a socio-political agenda can be interpreted from the collection. Pattern 8 in *A Pattern Language* compels the town planner to:

> Do everything possible to enrich the cultures and subcultures of the city, by breaking the city, as far as possible, into a vast mosaic of small and different subcultures, each with its own spatial territory, and each with the power to create its own distinct life style. Make sure that the subcultures are small enough, so that each person has access to the full variety of life styles in the subcultures near his own. (Alexander et al., 1977 p.42)

While we may perhaps agree with this claim on a personal level, it is hard to take it as an objective observation. As Erickson puts it: "Alexander's Pattern Language is not value neutral" (Ericksson, 2000). On the other hand, Alexander's Mexicalli project is taken as an emblem of participatory design, where patterns are used to facilitate design and empower users – who make their own choices (Dearden et al., 2002a). In this case, patterns are predominantly a social tool allowing the expert to communicate knowledge to the families designing their own home. One could claim that there is a socio-political agenda here as well. The difference is that in this case it is made explicit, and given as the premises – not the conclusion.

Such pattern languages seem to be quite alien to the *descriptive* pattern languages, prevalent in software design. While in urban planning and architecture it is clear that almost any decision has a political and ideological context, it is hard to see such context in the design of, for example, network routing protocols. However, the normative dimension of technology cannot be avoided in technology-enhanced learning. After all, everything about education is inherently value-driven. Every piece of technology designed for education assumes – and therefore supports – a particular organizational structure and a specific prioritization of knowledge. Yet these assumptions are often left unmentioned.

*Design patterns for learning*

The first reference to learning was made by (Alexander et al, 1978) when he described a pattern called "Network of Learning". The approach in education more generally has manifested itself through three main trends. The first is the growing trend of pedagogical design patterns (Anthony, 1996; Bergin 2000; Eckstein, Bergin & Sharp, 2002). The second is the development of software design patterns for educational technology (Dearden, et al., 2002b; Avgeriou et al, 2004). The third is the search for patterns in related practices, such as evaluation and assessment (Chaquet, El-Kechaï & Barre, 2005) and and analysis of learning and learning systems (Gibert-Darras et al., 2005).

Pedagogical design patterns apply the concept of design patterns to pedagogical design. The fundamental claim behind this effort is that many experienced practitioners in education have tried and tested methods of solving recurring problems or addressing common needs. Among the pioneers in this field were Anthony (1995) and later the pedagogical patterns project (http://www.pedagogicalpatterns.org/), initiated by a group of experienced software engineering and computer science educators (Bergin, 2000; Eckstein, Bergin & Sharp, 2002). They proposed a set of patterns dealing with issues ranging from the design of a college course to specific principles of computer science instruction and to concrete problems and their solutions.

A second arena that has seen a proliferation of design patterns is web-based educational technologies. Notable examples in this field include the E-LEN project (http://www2.tisip.no/E-LEN/) and several initiatives within the IMS-LD framework (http://www.imsglobal.org). Most of the work in this area is focused on the engineering aspects of designing, developing, deploying and evaluating good technology for web-based instruction (Frizell & Hubscher, 2002; Hernández-Leo et al, 2006; Bailey et al, 2006).

This strain of work is done mainly in the context of developing large scale technological systems to support organizational and vocational learning or web-delivered higher and further education. Due to this context, much of the work is highly technical. Many of the valuable innovations have a strong engineering flavour to them (e.g. Bailey et al, 2006) which might deter teachers and educational researchers. Even the issue of uncovering design patterns can get embellished as structural analysis of XML documents (Brouns et al, 2005). The interaction between student and instructor is assumed to be mediated by this communication channel. Under such circumstances, most of the effort goes into designing the representation and organization of educational content and the mechanisms by which learners interact with it (Frizell & Hubscher, 2002). Design patterns are also situated in this context, with the engineer of educational technologies as the user in mind (Avgeriou et al, 2003; Garzotto et al, 2004; Kolås & Staupe, 2004).

A related approach is offered by the design principles project (Kali et al, 2004; Kali, 2006; Linn & Eylon, 2006). Design principles are organized in three layers: meta-principles, pragmatic principles and specific principles. While the meta-principles are comparable to pedagogical foundations of our approach, the pragmatic principles bear resemblance to Alexanderian design patterns. While design patterns are densely linked at the language level, they are also self contained in the sense that they include the context and problem where they apply. In the design principles approach, this information is stored in the links between principles of different levels, and between principles and features.

In an attempt to distance himself from the dominant approaches in e-learning, Goodyear (2004) focuses on what he calls networked learning, where technology is used to promote connections between learners and foster communities which make efficient use of their resources. In this context, Goodyear emphasises patterns as a means of empowering practitioners to utilize accumulated design knowledge. His patterns are succinct and written in plain language.

Another study oriented towards educators is (Dearden et al, 2002a; 2002b). They point out the strong ideological and methodological parallels between Alexander's original vision of pattern language and the paradigm of participatory design. Pattern languages were conceived as a means of making expert knowledge accessible to naive planners, and enable educated and informed designers to work with naive users in collaboration. By contrast, in practice many pattern languages have taken a highly specialized form, and have become part of a professional jargon. As an alternative, Dearden et al propose the 'facilitation' model developed by Alexander et al (1985) in the Mexicali project. In that project, an 'Architect-builder' worked with a family to enable them to design and build their own house. Very significantly, the pattern language was shared by the designer and the family, and used to present and discuss design problems and solutions. The family could refer to the pattern even when choosing an alternative design.

Participatory design is one of the most exciting and challenging paradigms to emerge in educational research over the last decade. Participatory design is "a set of theories, practices, and studies related to end users as full participants in activities leading to software and hardware computer products and computer-based activities" (Muller and Kuhn, 2002). From this perspective, Béguin (2003) points at the close relationship between design and learning. He suggests that effective design should be constructed as a process of mutual learning involving users and designers and argues that the products only reach their final form through use. This should be reflected in an iterative design process which allows the users and designers to collaboratively shape their concept of the product and its actual form simultaneously. Such an approach, if sometimes not explicitly stated in these terms, has led to the emergence of methodologies, which utilize the participatory design of tools and artefacts as a central element in the learning process. For more, we refer the reader to the insightful reviews by Druin (2002) and Nesset and Large (2004), and the recent work by Kaptelinin, Danielsson and Hedestig (Kaptelinin et al, 2004; Danielsson 2004).

One of the studies Dearden at al. report uses Bergin's language pedagogical patterns to support the participatory design of an elearning web-site. The design was produced by a group of students and practicing teachers and facilitated by an experienced designer. They report that using this approach empowered the practitioners and enabled them to produce quality designs. This approach also enabled the facilitator to structure the design process and communicate complex issues. On the cautionary side, practitioners reported initial difficulties and even stress associated with learning such a new approach. They also tended not to question the patterns, relying on them as given truths. These issues place additional responsibility on the facilitator.

A major research challenge is to communicate the potential of tools developed in technology-oriented research to the pedagogy and epistemology research communities, and vice-versa. Design patterns have the potential to bridge between these disparate research and practice communities, and allow each one to enjoy the fruits of the other's efforts. In order to materialize this potential, pattern languages need to avoid jargon, and at the same time make space for higher theoretical discussion. They should be based on a theoretical layer concerning pedagogy and epistemology and consider the learning context.

At this point, it would befit to observe several examples of patterns. Aside from the space limitations, we are confronted with the problem of linearity: the densely connected nature of pattern languages presents a challenge when representing them on paper. In fact, it was this aspect which led Ward Cunningham to develop the first Wiki to host the Portland Pattern Repository. We therefore urge the reader to inspect and discuss the current state of our pattern language at: http://lp.noe-kaleidoscope.org.

## Discussion

Design approaches in technology-enhanced learning have a single unifying constraint: the learner. Every tool developed aims to enable the learner to achieve an insight into a particular subject and each design approach supports this underlying philosophy. Simon (1969) argues for rigorous standards through a science of design. This offers the opportunity for identifying, in a structured manner, the points at which the design (as it developed) achieves its intended aim. This is especially important when developing ILEs, as critical interactions for learners can be specified and built into the design. In Simon's words, design practice can then attempt to match the "inner" specification to the "outer" goals, i.e. the ILE specification for a learning task to the constraints of the learning context. This view of the role of design practice is supported by DiSessa and Cobb (2004), whose ontological innovation is aimed at "hypothesizing and developing explanatory constructs, new categories of things in the world that help explain how it works". Here, the role of theory must play a central part in generating design alternatives that are relevant to the learner and their practice of learning. However, this is not to say that the design process is linear and prescriptive. Indeed, they point out that evidence of interesting forms of learning and new lines of inquiry often occur opportunistically and retrospective analysis may often be required to validate them. The results will then provide a basis for follow-up work, leading to an 'ontological innovation', which can then itself be refined, testing and extended.

Design patterns, when considered as a language, are both a research and a practical tool. As a research tool, they can be developed to encapsulate researchers' knowledge in a form that is transferable and applicable to various learning contexts, and is accessible to practitioners as a pragmatic resource. In this sense, they have the potential to be used by researchers for undertaking complex TEL-related tasks, for example, as a support for developing ontological innovations. They can also be developed and used as a practical tool; for example, directly addressing TEL design issues with regard to technological development and tool deployment. Essentially, once patterns are available, anyone involved in the TEL development process can take them, in any variety of combinations, and use them to (collaboratively) design their own tools. Additionally, each pattern can be evaluated and modified to take into account the context of use. As the patterns form a language, they can be ordered beginning with patterns that detail high level actions and processes (for example, Storyboarding: Negotiating the initial design of a game between a teacher and a developer), right down to those that detail coding techniques (for example, Alter an object's behaviour when its state changes). Patterns lower down should build, in a bottom-up manner, to complete those above them (Alexander et al., 1977).

Some design patterns which have originally been conceived in response to pragmatic engineering issues may have potential pedagogical value, if put in the right context. Mor et al (2006) demonstrate the potential epistemic benefits of using a software design pattern called Streams in learning about number sequences. This claim is stated in the context of constructionist programming activities, in which children generate and manipulate number sequences. Traditionally in such contexts, a sequence would be represented by a list: a static array of the first *n* terms of the sequence. The Streams pattern replaces this structure with a dynamic process that generates the terms on the fly, and passes them from one module to the next, in a manner similar to a factory production line. Using this design pattern allowed students to mould their intuitions into a situated formalism with which they could explore quite complex ideas, and argue their hypotheses convincingly and with commitment.

Another example is the model-view-controller pattern (Gamma et al, 1995). This pattern dissects the representation and manipulation of information from its structure and content. Perhaps one of the most powerful patterns in interface design, it also resonates the pedagogical discussion of representations (Balacheff & Kaput, 1996; Radford, 2000). Indeed, this pattern is utilized in the design of the ToonTalk animated programming language (Kahn, 1996). However, it needs to be explicitly communicated to educators involved in constructionist activity design for them to leverage the extensive body of computer science knowledge it embodies. Such an effort should contextualize this pattern in both worlds: that of software engineering and that of educational design.

We pointed out that design patterns could be used to capture stages in the process of developing ontological innovations. As ontological innovations are developed to help explain how the world works, design patterns act as a construct by which to help select and validate the design alternatives generated. They have value at each stage from opportunistic research, retrospective analysis and discovery to refinement, iteration and extension. In particular understanding and sharing patterns that help researchers with retrospective analysis of data and with how to follow up on opportunistic events would be useful. By developing such patterns all participants in the process have the means by which to discuss and analyse the development of ontological innovations.

Two of the issues we noted earlier suggest open issues for future research. The first regards the role of design patterns in participatory design, and the second concerns the relationship between patterns and representation.

Dearden et al (2002a; 2002b) stand out in their declared agenda of using patterns in a participatory design process. In most cases, pattern language initiatives tend to end up offering a cookbook of expert recipes for success. This approach suffers from several weaknesses. First, it fails to acknowledge the learning process of the designers themselves. The cohesive community process of developing the pattern languages often produces a pattern jargon, impenetrable to novices. We argue that pattern language development should be accompanied by the development of participatory methodologies and the technologies to support them.

As we saw above, the choice of representations is a critical factor in the design of ILEs. Yet most of the design research in this area offers conclusions of the sort 'representation X worked for problem Y'. Such observations, as perceptive as they may be, lack predictive power. What we are looking for are statements of the form 'if the problem has feature Y, then you should consider representation X because…'. This structure is starting to resemble a design pattern. Existing pattern languages focus on the structure of tools and activities. We suggest that pedagogical design patterns should also offer a way of representing pedagogical knowledge in a form that is both practical and refutable.

Scaife et al (1997) hint at a direction which may integrate both issues we noted. While reporting on a study in the participatory design of an ILE, they discuss issues of representation – both with respect to artefacts used in the process, and in terms of the resulting system. Perhaps a design pattern approach could offer a powerful method for identifying representations suitable for learning, by allowing the learners to participate in the design of these representations.

## Conclusion

Understanding design approaches in technologies enhanced learning is fundamental to the ability of TEL researchers and practitioners to carry out their work. This is especially true in a world of fast-paced technological advances, where learners have come to expect a high degree of sophistication from the technologies they engage with. Moreover, the relationship between researchers, practitioners, learners and technological developers is becoming ever more critical. No longer can one community attempt to design TEL tools; communication and expertise sharing amongst them is of paramount concern.

A fundamental problem that TEL researchers as a community face is the lack of cumulatively built understandings within the field. In the worst case, this affects the ability of the community to develop key underlying theories and methodologies for solving many of the critical problems concerning the use of technologies in learning and education, or to find ways to apply theories coherently at the level of design. If we look to the natural sciences, there is a direct link between current and previous research, leading to well—founded cumulative knowledge. In the social sciences such linearity is difficult to achieve because, by their very nature, the social sciences are embedded with real—world complexities and contradictions: and worse still, they involve those unpredictable beings – people – who act back on the system. However, design approaches – and design patterns in particular – offer the potential to support cumulativity by providing a common language for researchers to communicate their research. The possibility of engendering a culture of cumulativity is within our grasp.

## Acknowledgements

We would like to thank Prof. Richard Noss and the paper reviewers for their insightful comments on this paper. This work was supported by Kaleidoscope Network of Excellence, supported by the European Community under the Information Society Technologies priority of the 6th Framework Programme.